\documentclass[letterpaper, 10 pt, conference]{ieeeconf}  

\IEEEoverridecommandlockouts                              
\usepackage{graphicx}
\usepackage{subfig}
\graphicspath{{figuras/}}
\usepackage{cite}

\title{\LARGE \bf
Particle Motion Analysis in the Bronnikov black hole solution
}

\author{Katharine I. Cuba$^{1*}$ and Santiago E. Perez$^{2*}$
\thanks{*This work was not supported by any organization}
\thanks{$^{1}$H. Katharine I. Cuba is with Faculty of Physics,
        Universidad Nacional de Trujillo, Trujillo, Per\'u.}%
\thanks{$^{2}$Santiago E. Perez is with Instituto de F\'isica Armando Dia Tavares, Universidade Estadual do Rio de Janeiro - UERJ, Rio de Janeiro, Brazil}%
}

\begin{document}

\maketitle
\thispagestyle{empty}
\pagestyle{empty}

\begin{abstract}

There is a growing interest in the study of regular black holes because they eliminate the problem of the singularity that conventional black hole models present. We analyze here the motion of different types of particles (without mass, massive and massive and charged particles) in an specific regular charged black hole solution obtained by Bronnikov in 2001. We also examine the geometry and the nature of the Bronnikov black hole, and compare it with the Reissner-Nordstrom black hole.

\end{abstract}

\section{INTRODUCTION}

The first solutions of the equations of General Relativity (GR) representing black holes (namely those of Schwarzschild, Reissner-Nordstrom and Kerr) have an important feature in common, the presence of singularities. Singularities represent a problem since the laws of physics do not work in ``there'' \footnote{See \cite{Romero2013}, \textit{Adversus singularitates: the ontology of space-time singularities} }. Hence the interest for regular black hole solutions
in which the singularity is absent. One possible form of source of Einstein's equations 
is 
leading to nonsingular solutions is
nonlinear electrodynamics (NLED). Starting from the work  by Bardeen (1968), many solutions of regular charged black holes with NLED as a source have been discovered (see for instance the work done by Ay\'on-Beato and Garc\'ia \cite{AyGa,AyGa2}, Cataldo and Garc\'ia \cite{CataGa}, Dymnikova \cite{Dymnikova}  and Novello \cite{NoPeSa2000}). In this contribution we have analyzed the solution presented by  Bronnikov (2001) \cite{Bronnikov2001}, where a certain Lagrangian 
${\cal L}={\cal L}(F)$ (where $F =F_{\mu\nu}F^{\mu\nu}$ and $F_{\mu\nu}$ is the Maxwell tensor) describing a NLED 
was shown to lead to
regular solutions with magnetic charge. 
Using this specific example of regular black hole, we study the motion of different kind of particles (charged particles, with mass and without mass) in it. The results are compared with those obtained for the Reissner-Nordstrom solution \cite{Reissner1916,Nordstrom1918}. We present first  the features of both black holes and analyze the effective potential for each kind of particle followed by the numerical integration of the orbit equation in some cases of interest.

\begin{figure*}[tbh]
 \centering
\centerline{\subfloat[]  {{\includegraphics[width=200pt]{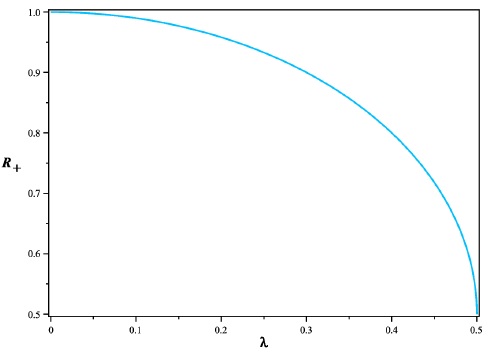}}} 
\hspace{1.75cm}
\subfloat[]{{\includegraphics[width=170pt]{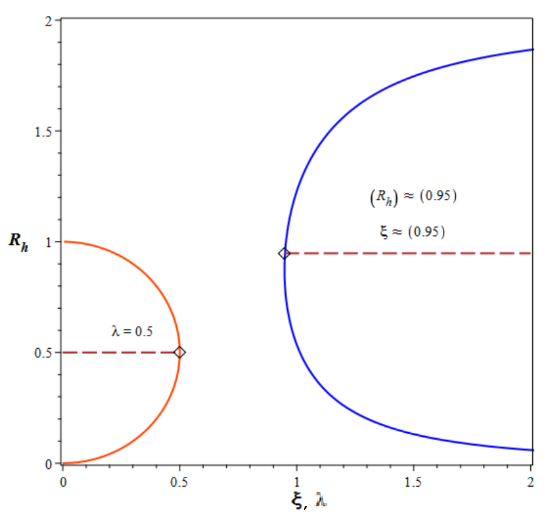}}} }
 \caption{(a) Shows the dimensionless outer horizon $R_+$ as a function of the dimensionless charges $\lambda=\frac{Q_e}{2M}$, we see that when $\lambda=0$ we obtain the Schwarzschild horizon and when $\lambda=0.5$ we obtain an extreme case for the RN black hole, that means that when the electric charge and the gravitational mass of the black holes has the same magnitude $Q_e=M$. In (b) we see the graph of horizons for both cases, orange line represent the horizons for the RN black hole and the blue line represents the horizons for the Bronnikov black hole, in function of its respective dimensionless parameters $\lambda=\frac{Q_e}{2M}$,$\xi=\frac{M_EM}{|Q_m|}$.
  As we see both cases have an extreme value in $\lambda=0.5$ and $\xi=0.5$ and, however the Bronnikov black holes is not limited for the values its parameter $\xi$ takes (after its minimum in the extreme value) and admits magnetic charges $Q_{m}$ bigger its electromagnetic mass $M_{EM}$.}
\label{RNhorizon}
\end{figure*}

\section{CHARGED BLACK HOLES}

\subsection{Bronnikov's Black Hole}

The action of the General relativity coupled with an arbitrary electromagnetic theory is given by:
\begin{eqnarray}
S=\frac{1}{16\pi}\int d^{4}x\sqrt{-g}[R-{\cal L}(F)],
\end{eqnarray}
Where $R$ is the Ricci scalar, $F_{\mu\nu}=\partial_{\mu}A_{\nu}-\partial_{\nu}A_{\mu}$ is the electromagnetic field tensor, $A_{\mu}$ is the four-potential, 
and ${\cal L}$ is an arbitrary function such that ${\cal L}(F)\approx F$ when $F\rightarrow 0$. The tensor $F_{\mu\nu}$ obeys the 
equations:
\begin{eqnarray}
\nabla_{\mu}({\cal L}_{F}F^{\mu\nu})=0, && \nabla_{\mu}^{*}F^{\mu\nu}=0.
\label{eqn:bianchi}
\end{eqnarray}
Here $^{*}F^{\mu\nu}=\epsilon_{\mu\nu\sigma\lambda}F^{\sigma\lambda}$ 
and ${\cal L}_{F}=\frac{d\cal L}{dF}$.
For spherical symmetry we just have the radial components of $F_{\mu\nu}$ corresponding to the electric and magnetic fields.
The energy momentum tensor that follows from the above given action is
$T^{\nu}_{\mu}=2{\cal L}_{F}F_{\mu\alpha}F^{\nu\alpha}-\frac{1}{2}\delta^{\nu}_{\mu}{\cal L}.$
It is useful to define the next expressions
$f_{e}=2F_{01}F^{10}$ and
$f_{m}=2F_{23}F^{32}$,
such that that energy-momentum tensor takes the form (see \cite{Bronnikov2001} pages 1 and 2)
$T_{\nu}^{\mu}=\frac{1}{2}{\rm diag}[2f_{e}{\cal L}_{F}+{\cal L},2f_{e}{\cal L}_{F}+ 
{\cal L},\nonumber \\ 
2f_{m}{\cal L}_{F}+{\cal L},2f_{m}{\cal L}_{F}+{\cal L}]$.
The line element corresponding to a static spherically symmetric space-time is given in the general form by:
\begin{eqnarray}
ds^{2}=-e^{2\gamma(r)}dt^{2}+e^{2\alpha(r)}dr^{2}+r^{2}d\Omega^{2},
\end{eqnarray}

Where $d\Omega^{2}=d\theta^{2}+\sin{\theta}^{2}d\phi^{2}$, and $r$ is the radial coordinate. Due to the spherical symmetry, the Maxwell's tensor just take the radial electric and radial magnetic field, it follows from equation (\ref{eqn:bianchi}) that
$r^{2}e^{\alpha+\gamma}L_{F}F^{10}=Q_{e}$ and
$F_{23}=Q_{m}\sin{\theta}$,
where $Q_{e}$ is the electric charge and $Q_{m}$ is the magnetic charge.
From Einstein's equations $G^{\mu}_{\nu}=-T^{\mu}_{\nu}$ and considering the energy-momentum tensor 
we get the line element in the form,
\begin{eqnarray}
ds^{2}=-\left(1-\frac{2M(r)}{r}\right)dt^{2}+\left(1-\frac{2M(r)}{r}\right)^{-1}dr^{2} \nonumber \\
+r^{2}d\Omega^{2},
\label{eqn:line element}
\end{eqnarray}
where
\begin{eqnarray}
M(r)=\frac{1}{2}\int r^{2}T^{0}_{0}dr + k.
\label{eqn:M function}
\end{eqnarray}
In the case of vacuum, 
$T^{0}_{0}=0$, and the solution reduces to that of the 
the Schwarzschild black hole, with $k=2M$, where $M$ is the gravitational mass \footnote{All this mathematical derivation was made by Bronnikov in \cite{Bronnikov2001} }. We will use next the results presented here to study the Bronnikov black hole solution and 
compare some of its features with those of the Reissner-Nordstrom black hole. 

\begin{figure*}[tbh]
 \centering
\centerline{\subfloat[Effective potentials of the Reissner- Nordstrom and Bronnikov geometries for particle without mass.]{{\includegraphics[width=140pt]{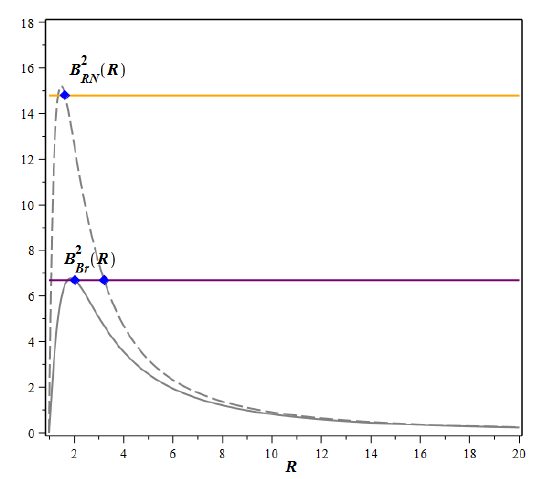}}} \hspace{1.4cm}
\subfloat[ Effective potentials of the Reissner- Nordstrom and Bronnikov geometries for massive particle.]{{
\includegraphics[width=130pt]{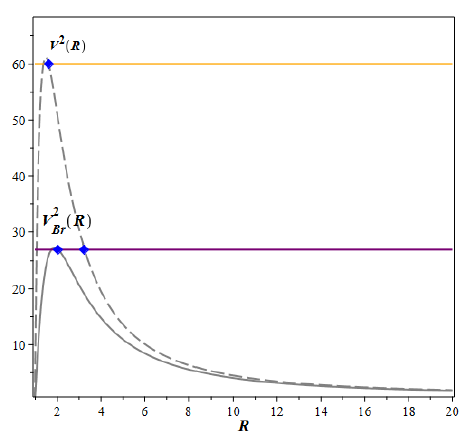}}}
\hspace{1.4cm}
\subfloat[ Effective potentials of the Reissner- Nordstrom and Bronnikov geometries for massive particle and charged particles.]{{
\includegraphics[width=135pt]{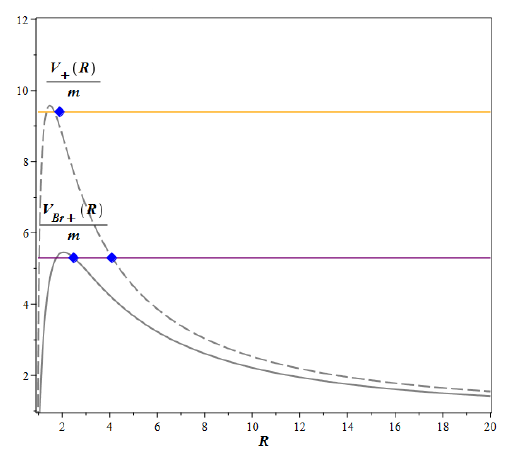}}} }
\centerline{\subfloat[ Orbits described for particles without mass in the two geometries.]{{
\includegraphics[width=140pt]{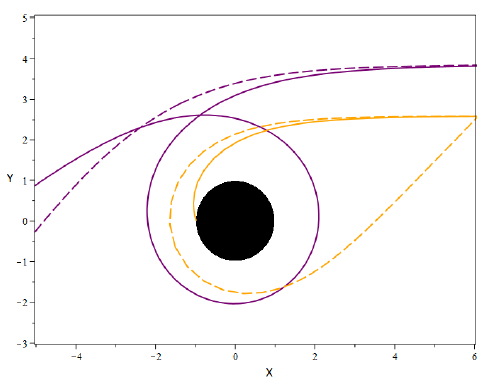}}}\hspace{1.4cm}
\subfloat[ Orbits described for massive particles in the two geometries. ]{{
\includegraphics[width=130pt]{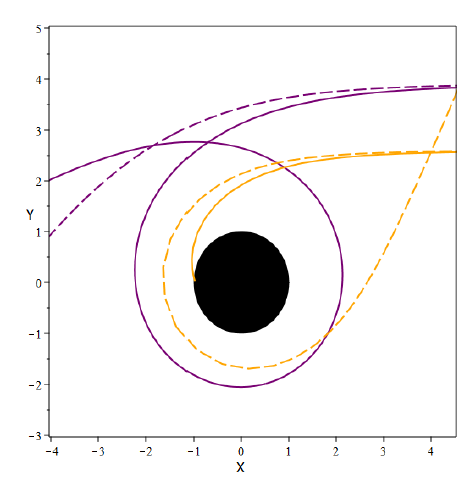}}}
\hspace{1.4cm}
\subfloat[(Orbits described for massive and charged particles in the two geometries.]{{
\includegraphics[width=140pt]{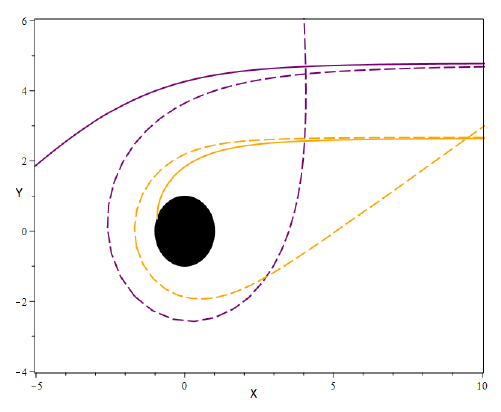}}}
 }
 \caption{Effective potential and orbits described for the three types of particles studied in the Reissner-Nordstrom and Bronnikov geometry. In figure (a) we see the effective potentials of the two geometries for particles without mass and its corresponding trajectories in figure (c), the angular momentum for both cases is $L_{\bullet}=\tilde{L}=20$ and the values of energy chosen are $E=6.7$ and $E=14.8$. Figure (b) shows the effective potentials for massive particles and its corresponding graph of the trajectories made by massive particles in (d), both cases have an angular momentum of $L_{\star}=20$, $E_{*}=27$ and $E_{*}=60$. Figure (c) shows the effective potentials for the case of massive particles with charge in the Reisser-Nordtrom and Bronnikov black hole, in (c) we see the orbits made for this type of particle with an angular momentum $L_{\star}=25$ with energies $E_{*}=5.3$ and $E_{*}=9.4$.  In the three cases solid lines represents the potentials and trajectories in the Bronnikov black hole and dashed lines represent the potentials and trajectories in the Reissner-Nordstrom black hole. } 
 \label{OrbitsRnBr}
\end{figure*}

\subsection{Reissner-Nordstrom Black Hole}

The metric of the Reissner-Nordstrom black hole is: $g_{RN}(r)=\left(1+\frac{2M}{r}+\frac{Q_{e}^{2}}{r^{2}} \right)$. The position of the horizons follows from $g_{RN}(r)=0$, and is given by  $r_{\pm}= M\pm \sqrt{M^{2}-Q_{e}^{2}}$. Here $r_{+}$ is the outer horizon and $r_{-}$ is the inner horizon. From now on, we shall use the dimensionless coordinate $R_{\pm}=\frac{r\pm}{2M}=\frac{1}{2}+\sqrt{\frac{1}{4}-\lambda^{2}}$ where $\lambda=\frac{Q_{e}}{2M}$.

\subsection{Bronnikov's Black Hole}

The necessary and sufficient conditions for the existence of static black holes solutions with spherical symmetry of Einstein's equations with NLED as a source (with the Maxwell limit for weak fields) were discussed by Bronnikov (2001). The result is that the only case in which a regular solution can be obtained is $Q_{e}=0$ and $Q_{m}\neq 0$. In this case 
$T^{0}_{0}=\frac{1}{2}{\cal L}(F)$,
and from  (\ref{eqn:M function}) we get 
$M(r)=\frac{1}{4}\int {\cal L}(F)r^{2}dr +k$.
Bronnikov (2001) chose $k=0$, since a nonzero value would introduce a $1/r$ term in $M(r)$, thus reintroducing the singularity.  He also worked with a specific for of the Lagrangian given by:
\begin{eqnarray}
{\cal L}(F)=\frac{F}{\cosh^{2}\left({a|\frac{F}{2}|^{1/4}}\right)},
\label{eqn:Bronnikov L function}
\end{eqnarray}
where $a$ is a constant and $F=\frac{2Q_{m}^{2}}{r^{4}}$. Substituting equation (\ref{eqn:Bronnikov L function}) in the equation $M(r)$ 
we have:
\begin{eqnarray}
M(r)=\frac{Q_{m}^{3/2}}{2a}\left[1-\tanh{\left(\frac{a\sqrt{|Q_{m}|}}{r} \right)} \right].
\end{eqnarray}

The Bronnikov regular black hole geometry is given for the line element
$ds^{2}=-g_{BR}(r)dt^{2}+g_{BR}^{-1}(r)dr^{2}+r^{2}d\Omega^{2}$,
with $g_{BR}=1-2M(r)/r$ given for the expression:
\begin{eqnarray}
g_{BR}(r)=\left(1- \frac{Q_{m}^{3/2}}{2a}\left[1-\tanh{\left(\frac{a\sqrt{|Q_{m}|}}{r} \right)} \right] \right).
\end{eqnarray}
Notice that $g_{BR}(0)=1$, and that the Ricci scalar is given by the expression,
\begin{eqnarray}
 R=\{-2Q_{m}^{2}a \Bigl[\sinh{\left(\frac{a\sqrt{q_{m}}}{r} \right)}\cosh{\left(\frac{a\sqrt{q_{m}}}{r} \right)}\times \nonumber \\
 \left(a\sqrt{Q_{m}}+\frac{Q_{m}^{2}}{r} \right) +\frac{Q_{m}^{2}}{r}\sinh^{2}{\left(\frac{a\sqrt{Q_{m}}}{r}\right)}\Bigl]\}\times \nonumber \\
 \{ r^{5}\cosh^{3}{\left(\frac{a\sqrt{Q_{m}}}{r} \right)}\Bigl[\frac{Q_{3}^{}3/2}{r}\sinh{\left(\frac{a\sqrt{Q_{m}}}{r} \right)}+ \nonumber \\
 \cosh{\left(\frac{a\sqrt{q_{m}}}{r} \right)}\left(a-\frac{Q_{m}^{3/2}}{r} \right)   \Bigl]\}^{-1}.
\end{eqnarray}

Evaluating the Ricci scalar at $r=0$ we have that $\lim_{r\to 0}{R}=0$, in agreement with the regularity of the solution. In fact, the calculation of the components of the Riemann tensor in tetrads shows that the solution is regular in $r=0$. It is important to note that the total mass associated with this solution is given by 
$M_{EM}=M(\infty) = \frac{|Q_{m}|^{3/2}}{2a} $ (See \cite{Manko2016,Pellicer1969}). The subindex ``EM" indicates that the total mass is purely electromagnetic, since $M_{EM}=0$ for $Q_{m}=0$. It is convenient to rewrite
$g_{Br}(r)$ in terms of
the parameter $\xi=\frac{M_{EM}}{|Q_{m}|}$ or $\xi=\frac{|Q_{m}|^{1/2}}{2a}$ 
to determine the horizons of  Bronnikov's black hole. The result is:

\begin{eqnarray}
\xi=\left[R_{h}\ln{\left(\frac{4-R_{h}}{R_{h}}\right)} \right]^{1/2},
\end{eqnarray}
where $R_{h}=\frac{2a r_{h}}{|Q_{m}|^{3/2}}$ and $r_{h}$ determines the position of the horizons. 
Figure \ref{RNhorizon}(b) shows the curves corresponding to the Reissner-Nordstrom horizons $(R_{\pm})$ and the Bronnikov horizons $(R_{h})$. Here we see that 
in the case of the RN black hole,
$R_{-}$ and $R_{+}$ are limited below and above by $0$ and $1$ respectively, and the extremal case corresponds to $\lambda =0.5$. 
for Bronnikov's solution, the extreme case corresponds to $\xi=\xi_{0}\approx 0.95$, but $R_+$ can grow without limit.

\section{ORBITS}

In this section, the motion of  massless particles, as well as  massive neutral and charged particles, is studied in  Bronnikov's solution, and is compared to that in the RN gometry. We start with the analysis of the effective potential, and then we numerically calculate the orbit for each kind of particle, in some cases of interest.  We shall adopt below the following expression for the metric coefficient 
$g_{Br}(R)=\left(1-\frac{8\kappa\xi^{3}}{R}\left[1-\tanh{\left( \frac{2\kappa\xi}{R}\right)} \right]\right)$,
where $R=\frac{r}{M_{EM}}$ and $\kappa^{2}=\frac{a^{2}}{M_{EM}}$.

\subsection{Particles without mass $m=0$} 
The effective potential for particles without mass (with the exception of photons) \footnote{In nonlinear theories of electromagnetism, photons follow trajectories determined by an effective metric. See Goulart de Oliveira, E. and Perez, S.: \textit{A classification of the effective metric in nonlinear electrodynamics.}} is expressed as
\begin{eqnarray}
B^{2}_{Br}(R)=g_{Br}(R)\frac{L^{2}_{\bullet}}{R^{2}},
\end{eqnarray}
where $L_{\bullet}=\frac{L}{M_{EM}}$, $R=\frac{r}{M_{EM}}$ and $L$ is the angular momentum of the massless particle.

The equation of the orbits for massless particles is:
\begin{eqnarray}
\phi_{Br}(R)=\int_{R'}^{\infty}\left[\frac{R^{4}}{b^{2}_{Br}}-g_{Br}(R)R^{4} \right]^{-1/2}dR,
\end{eqnarray}
where $b_{Br}=\frac{L_{\bullet}}{E}$ 
and $E$ is the energy. $R'$ is the turning point of the motion, or the outer horizon radius. 

\subsection{Particles without mass $m\neq0$ and $q=0$} 
In this case, the effective potential is given by the equation,
\begin{eqnarray}
V^{2}_{Br}(R)=g_{Br}(R)\left(1+\frac{L_{\star}^{2}}{R^{2}}  \right),
\end{eqnarray}
where $L_{\star}=\frac{L}{mM_{EM}}$. As a consequence, there is a critical value given for $L_{\star c}\approx 9.19$ such that for $L_{\star}>L_{\star c}$ the potential has two extreme points, which do not happen in the RN case.

The equation of the orbit for particles with mass is:
\begin{eqnarray}
\phi_{Br}(R)=\int_{R'}^{\infty}\left[\frac{R^{4}E_{*}^{2}}{L_{\star}^{2}}-
g_{Br}(R)\left(\frac{R^{4}}{L_{\star}^{2}}+R^{2} \right)\right]^{-1/2}dR.
\end{eqnarray}
Here $E_{\star}=\frac{E}{m}$ and $R'$ represents also the turning point.

 
 \subsection{Particles with $m\neq0$ and $q\neq0$}

In this case, the effective potential (see Hobson \cite{Hobson2006}) is:
\begin{eqnarray}
\frac{V_{Br\pm}(R)}{m}= - \frac{\epsilon\xi}{R}\pm \sqrt{g_{Br}(R)\left(1+\frac{L_{\star}^{2}}{R^{2}}\right)},
\end{eqnarray}
Where $\epsilon=\frac{q}{m}$. Here $V_{Br\pm}$ represent the potential function. 
This potential is similar to that of the Reissner-Nordstrom black hole. 
However, there are some quantitative differences that modify the trajectories as explained later. 

The equation of orbit for this case is,
\begin{eqnarray}
\phi_{Br}(R)=\int_{R'}^{\infty}\Big[\left(\frac{R^{2}E_{*}}{L_{\star}}-\frac{\epsilon\xi R}{L_{\star}}\right)^{2}- \nonumber \\
g_{Br}(R)\left(\frac{R^{4}}{L_{\star}^{2}}+R^{2} \right)\Big]^{-1/2}dR.
\end{eqnarray}

If we plot the orbits described for this kind of particles we see that the larger the parameter $\xi$ is, the more closed is the orbit described by the particle. Remembering that the parameter $\xi$ depends of the magnetic charge, we can say that the more charged the black hole is, the greater its attraction to the particle will be. It is also interesting to note, from the relation $M_{EM}=\frac{|Q_{m}|^{3/2}}{2a}$, we have that $M_{EM}=Q_{m}$ for $Q_{m}=4a^{2}$ or $\xi=1$. This means that the Bronnikov solution admits the possibility to have a mass (of electromagnetic origin) smaller than the (magnetic) charge.  This in contrast with the Reissner-Nordstrom solution, where the (gravitational) mass  cannot be smaller than the (electric) charge.

\section{Comparison between the orbits of Reissner-Nordstrom solution and Bronnikov solution}

In this section we compare the characteristics of the motion of the different types of particles in the two geometries under scrutiny (Reissner-Nordstrom and Bronnikov). To achieve this goal, a common value for the external horizon $R_{+}$ has been taken for both geometries $R_{+}=0.98$. Such a value correspond to the parameter $\lambda=0.14$ (in the case of Reissner-Nordstrom) and $\xi\approx 0.952$ (in the case of Bronnikov). As we see in figure \ref{OrbitsRnBr}, in the three cases the attraction felt by 
all the three types of particles
due to 
Bronnikov's black hole 
is stronger that that due to 
he Reissner-Nordstrom black hole. This result is not limited to the values of energy and parameters examined here, we have checked that this behaviour is repeated for a large interval of the relevant parameters. 


\section{RESULTS, DISCUSSION AND PERSPECTIVES}

We have studied here the motion of particles (without mass, massive and massive and charged particles) in the Bronnikov solution and compared it with the Reissner-Nordstrom solution,
as well as some features of both solutions, we have shown that the position of the horizons from the Reissner Nordstrom black hole is always smaller than that of the horizons of the Bronnikov black hole. 

Regarding the comparison between the motion of particles in the two geometries, it was shown that the Bronnikov black hole
exerts a more intense attraction on the particles (regardless of the nature of the particle studied here) than the Reissner-Nordtrom black hole. Such a behavior may be due to difference in the behaviour of their energy density $(T_{0}^{0})$. 

We have also shown that there are important differences in the case of Bronnikov black hole with respect to Reissner-Nordstrom black hole, we see this principally in the relation "charge-mass". In the case of Reissner-Nordstrom solution, the parameter "$\lambda$" ($0\leq\lambda=\frac{Q_{e}}{2M}\leq 1$)
does not support that the electric charge of the black hole is greater than its gravitational mass, however for Bronnikov black hole, the parameter "$\xi$" ($\xi=\frac{M_{EM}}{|Q_{m}|}>\approx0.95$)
does support that the electromagnetic mass is bigger than its magnetic charge.

This work is just a small sample of the study of a specific case of a regular solution, research in this area is still in progress.



\end{document}